\documentclass{Interspeech}

\newcommand{\mmedit}[1]{\textcolor{black}{#1}} %

\usepackage{multirow}
\usepackage{amsmath,graphicx}
\usepackage[subrefformat=parens]{subcaption}

\interspeechcameraready

\title{Eigenvoice Synthesis based on Model Editing for Speaker Generation}
\author[affiliation={1,2}]{Masato}{Murata}
\author[affiliation=1]{Koichi}{Miyazaki}
\author[affiliation=1]{Tomoki}{Koriyama}
\author[affiliation=2]{Tomoki}{Toda}

\affiliation{}{CyberAgent}{Japan}
\affiliation{}{Nagoya University}{Japan}
\email{murata\_masato@cyberagent.co.jp, miyazaki\_koichi\_xa@cyberagent.co.jp, koriyama\_tomoki@cyberagent.co.jp, 
tomoki@icts.nagoya-u.ac.jp}

\usepackage{comment}

\begin{document}

\maketitle
\begin{abstract}

\looseness=-1
Speaker generation task aims to create unseen speaker voice without reference speech. The key to the task is defining a speaker space that represents diverse speakers to determine the generated speaker trait. However, the effective way to define this speaker space remains unclear. Eigenvoice synthesis is one of the promising approaches in the traditional parametric synthesis framework, such as HMM-based methods, which define a low-dimensional speaker space using pre-stored speaker features. This study proposes a novel DNN-based eigenvoice synthesis method via model editing. Unlike prior methods, our method defines a speaker space in the DNN model parameter space. By directly sampling new DNN model parameters in this space, we can create diverse speaker voices. Experimental results showed the capability of our method to generate diverse speakers’ speech. Moreover, we discovered a gender-dominant axis in the created speaker space, indicating the potential to control speaker attributes.

\end{abstract}

\section{Introduction}

Text-to-speech (TTS) technology has become widely used in various applications such as character voice creation in video content, conversational assistants, and audiobook narration. 
Due to the demand to generate voices \mmedit{with diverse speech expressions,} 
various studies have tackled related tasks, specifically
focused on generating speech with diverse speaker characteristics. 
Since traditional TTS models rely on the training datasets, they can only generate speech of existing seen speakers in the dataset but cannot generate that of unseen speakers. 
To address this limitation, zero-shot TTS has been proposed to generate speech of unseen speakers using a reference speech of the target speaker. 
However, this approach requires a reference speech, which \mmedit{is not always available. In other words, it has difficulty in generating ``really unseen'' speakers' voices.}
In contrast, there is a task called ``speaker generation'' that aims to generate diverse voices of unseen speakers without requiring reference speech~\cite{tacospawn}.
In the speaker generation task, defining a speaker space that represents diverse speakers is necessary to determine the speaker characteristics to be generated.
However, the effective way to define this speaker space remains unclear, and several prior studies have explored various approaches.

Eigenvoice synthesis~\cite{eigenvoice_synthesis} is one of the approaches for the speaker generation task that defines a low-dimensional speaker space using pre-stored speaker-dependent representations.
\mmedit{Then, this speaker space is spanned by basis vectors extracted from speaker-dependent representations of multiple pre-stored speakers (called base speakers).}
In this created speaker space, arbitrary speaker characteristics are represented as the weighted sum form of basis vectors.
These basis vectors are called ``eigenvoice'', which are mutually orthogonal and represent the \mmedit{dominant component among base speakers.}
This eigenvoice method~\cite{eigenvoice1, eigenvoice2, eigenvoice_survey} has been applied to TTS~\cite{eigenvoice_synthesis} and voice conversion~\cite{eigenvoice_conversion} tasks.
Previous studies succeeded in generating diverse speech characteristics by changing the weight coefficients of the basis vector.
Furthermore, this eigenvoice approach is applicable not only to the speaker generation task, but also to speaker adaptation and speech style control along \mmedit{the dominant axis.}

\mmedit{Although the eigenvoice methods are based on traditional probabilistic models, such as Hidden Markov Model (HMM), there are several other deep neural network (DNN)-based approaches based on a similar idea of creating a speaker space.}
Speaker embedding-based methods generate unseen speaker by sampling speaker embeddings from speaker distribution~\cite{tacospawn, mid_attribute}.
Text prompt-based methods represent speaker characteristics using text prompts~\cite{promptspeaker, promptttspp, generating_speaker}.
\mmedit{However, since these works rely on additional modules designed for the task, complex training processes or extensive datasets are required.}

\mmedit{In contrast,} recent research proposed a model merging speaker interpolation method that does not require any specifically designed additional modules~\cite{merge_attribute}.
Model merging is one of the model editing methods that interpolates the model parameters of two base models~\cite{model_soup}.
They reported that applying the model merging to TTS models can control speaker attributes.

By extending this model merging-based method, we propose a novel DNN-based eigenvoice synthesis method based on model editing.
Although the prior method~\cite{merge_attribute} represents the generated speaker characteristics by simply interpolating between two base models, 
our proposed method represents arbitrary speaker characteristics in a low-dimensional linear space in the high-dimensional DNN model parameter space.
Our method defines this linear space as ``speaker space''.
We can obtain new speakers' TTS model parameters by directly sampling from this speaker space, which achieves diverse speaker voice generation.
We evaluated the generated speakers' voices
regarding naturalness, intelligibility, and speaker diversity.
The results showed that our proposed method can generate diverse speakers while keeping speech quality.
Moreover, our analysis suggests the presence of a gender-dominant axis in the defined speaker space, indicating the potential for controlling gender attributes.
Audio samples are available on our demo page\footnote{\url{https://muramasa2.github.io/model-editing-eigenvoice/}}.

The main contributions of this paper are as follows:
\begin{itemize}
\item To the best of our knowledge, our method is the first DNN-based eigenvoice synthesis method that represents diverse speaker traits in high-dimensional model parameter space.
\item We demonstrate that our proposed method can synthesize diverse speakers' voices without using external modules.
\item Our experimental results suggest that the speaker space created by the proposed method has a gender-dominant axis, potentially enabling attribute manipulation.
\end{itemize}

\section{Model merging speaker interpolation based on task arithmetic}

Task Arithmetic is one of the model editing methods that directly edits trained model parameters using a task vector, which is defined as the difference between pre-trained model parameters and fine-tuned model parameters~\cite{task_arithmetic}.
Each task vector is considered to include the abilities for the target task.
By scaling and combining task vectors through various arithmetical operations, it creates new model parameters as follows:
\begin{align}
\label{equation:model_merge}
\boldsymbol{\tau} &= \boldsymbol{\theta}_{\mathtt{ft}} - \boldsymbol{\theta}_{\mathtt{pre}}
 \\
\boldsymbol{\theta}_{\mathtt{new}} &= \boldsymbol{\theta}_{\mathtt{pre}} + \alpha \cdot \boldsymbol{\tau}
\end{align}
where $\boldsymbol{\tau}$ is the task vector between a fine-tuned model $\boldsymbol{\theta}_{\mathtt{ft}}$ and a pre-trained model $\boldsymbol{\theta}_{\mathtt{pre}}$, and 
$\boldsymbol{\theta}_{\mathtt{new}}$ is the resultant new model parameters created with this task vector.
The resultant model is reported to obtain different model behaviors, such as performance improvements in image classification and speech recognition tasks~\cite{model_soup, merge_asr}, the acquisition of multiple language-related task capabilities~\cite{task_arithmetic, ties_merge}, and obtaining intermediate attribute outputs in image and speech synthesis~\cite{gan_cocktail, merge_attribute}.
Unlike conventional transfer learning, it does not require additional training.

\mmedit{Based on the task arithmetic method,} model merging speaker interpolation method~\cite{merge_attribute} was proposed for speaker generation task.
The resultant model parameters by merging single-speaker models A and B are defined as follows:
\begin{align}
\boldsymbol{\theta}_\alpha &= (1-\alpha) \boldsymbol{\theta}_\mathtt{A} + \alpha \boldsymbol{\theta}_\mathtt{B} \\
&= \boldsymbol{\theta}_{\mathtt{pre}} + \{(1-\alpha) \boldsymbol{\tau}_\mathtt{A} + \alpha \boldsymbol{\tau}_\mathtt{B} \}
\end{align}
where $\alpha$ is the merging coefficient, $\boldsymbol{\theta}_\mathtt{A}$ and $\boldsymbol{\theta}_\mathtt{B}$ are the existing fine-tuned weight parameters of each base TTS model, \mmedit{and $\boldsymbol{\tau}_\mathtt{A}$ and $\boldsymbol{\tau}_\mathtt{B}$ are the task vectors for each base TTS model.}

\section{Proposed method}
\label{section:proposed_method}

Inspired by
the model merging-based speaker interpolation method~\cite{merge_attribute}, we propose a novel eigenvoice synthesis method based on model editing.
\mmedit{Instead of the interpolating between two base models, our proposed method represents arbitrary speaker characteristics in a low-dimensional linear space in the DNN model parameter space.
While the prior eigenvoice method~\cite{eigenvoice_synthesis} used the stack of the HMM model parameters for base speakers, we utilize task vectors as speaker representations to obtain the basis vectors of the speaker space. 
By sampling from this speaker space, we can obtain diverse speakers' TTS model parameters, which allows us to generate diverse speaker voices.}

We first prepare $N$ single-speaker TTS models as the base models ($N$ is on the order of $10^1$ to $10^4$), each fine-tuned from the same pre-trained initialization.
Subsequently, we extract task vectors for each speaker (the updated value of the model parameter from the pre-trained model) and \mmedit{apply 1D flatten to it into $M$-dim vector, denoted as $\mathbf{a}_i \in \mathbb{R}^{M \times 1}, (i=1, \dots, N)$.
Then, $M$ denotes the number of the model parameters updated during fine-tuning process ($M$ is on the order of $10^5$ to $10^{10}$).
The $M$ value is extremely large compared to the number of base speakers $N$ (i.e., $M \gg N$).
In this study, this $M$-dim vector $\mathbf{a}_i$ is referred to as the speaker parameter vector.}
By stacking $N$ speaker parameter vectors as column vectors, we create a speaker matrix $\mathbf{A} = [\mathbf{a}_1, \ldots, \mathbf{a}_N] \in \mathbb{R}^{M \times N}$.
Each column vector $\mathbf{a}_i$ is considered to be a representation of the speaker characteristics of the $i$-th speaker.
We then apply singular value decomposition (SVD) to this speaker matrix $\mathbf{A}$, as follows:
\begin{align}
\mathbf{A} = \mathbf{U} \mathbf{\Sigma} \mathbf{V}^\top
\end{align}
\mmedit{where $\mathbf{U} \in \mathbb{R}^{M \times N}$ is a matrix whose columns are the left singular vectors of $\mathbf{A}$, $\mathbf{\Sigma} \in \mathbb{R}^{N \times N} $ is a diagonal matrix containing singular values of $\mathbf{A}$, and $\mathbf{V} \in \mathbb{R}^{N \times N}$ is a matrix whose columns are the right singular vectors of $\mathbf{A}$.
Now we define the column vector of $\mathbf{V}^\top$ as a speaker coefficient vector $\mathbf{w}_i , (i=1, \dots, N)$ as follows:
\begin{align}
\label{eq:speaker_modeling}
\mathbf{V}^\top = [\mathbf{w}_1, \dots, \mathbf{w}_N].
\end{align}
Then, each speaker parameter vector $\mathbf{a}_i$ can be expressed as follows:
\begin{align}
\label{eq:speaker_modeling}
\mathbf{a}_i = \mathbf{U} \mathbf{\Sigma} \mathbf{w}_i = \mathbf{B} \mathbf{w}_i
\end{align}
}where $\mathbf{B} \in \mathbb{R}^{M \times N}$ represents the basis vectors and $\mathbf{w}_i$ is the speaker coefficient vector corresponding to the $i$-th speaker.
This indicates that each base speaker parameter vector $\mathbf{a}_i$ can be represented as a weighted sum form of the basis vectors $\mathbf{B}$ with the speaker coefficient vector $\mathbf{w}_i$.

\begin{figure}[t]
   \centering
    \begin{minipage}[b]{1\linewidth}
      \centering
      \includegraphics[width=\hsize]{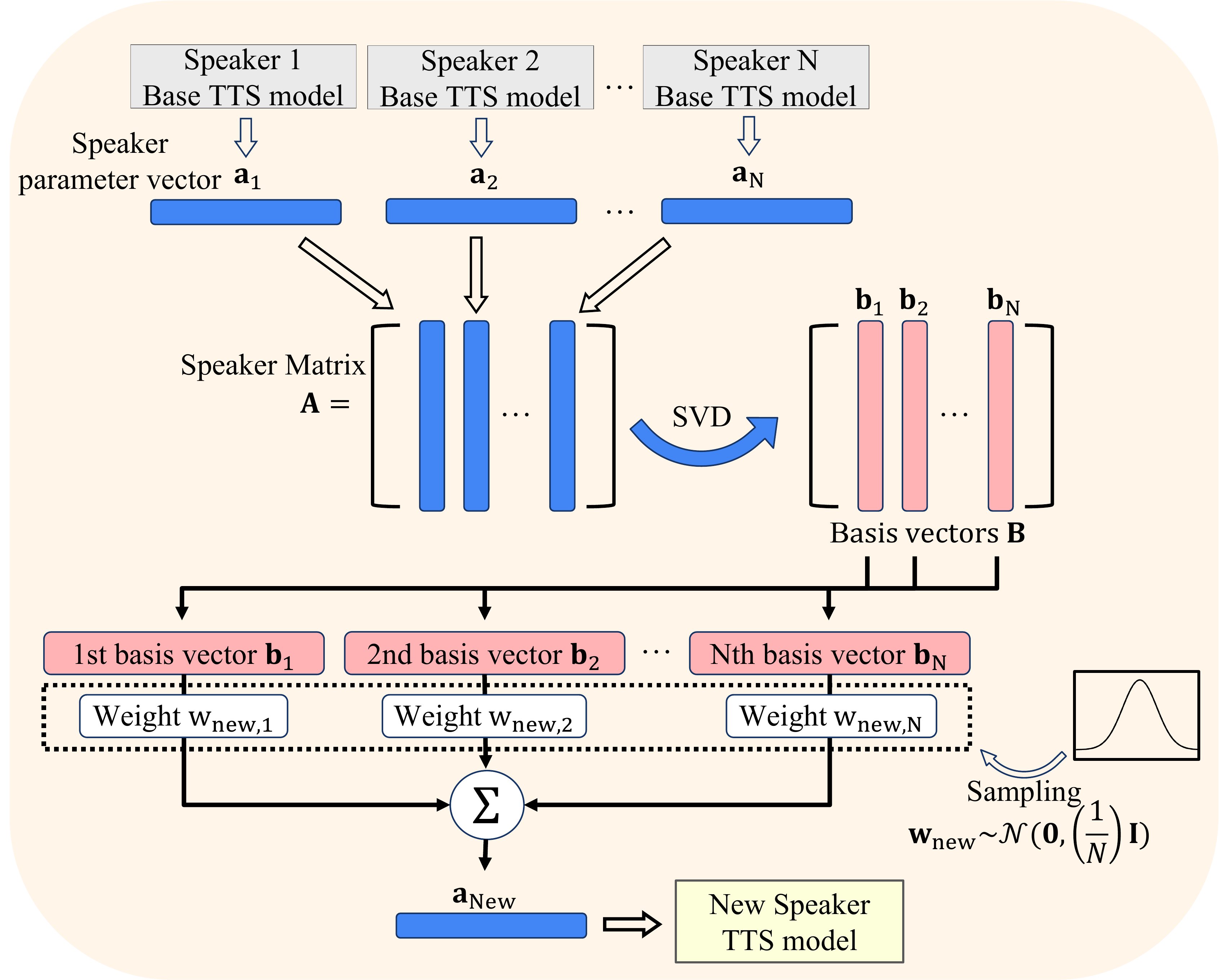}
    \end{minipage}
        \vspace{-15pt} %
    \caption{The process of generating the new speakers' TTS model by the proposed method. \mmedit{The blue and red rectangles show the speaker parameter vectors and basis vectors of the speaker space. The white rectangle shows the elements of the randomly sampled speaker coefficient vector.}}
    \vspace{-10pt} %
    \label{fig:t_sne}
  \end{figure}

\subsection{Speaker space in model parameter space}
\label{subsection:speaker_space}
As explained above, each base speaker parameter vector $\mathbf{a}_i$ can be represented as a weighted sum form of the basis vectors $\mathbf{B}$ with the speaker coefficient vector $\mathbf{w}_i$.
Hence, the speaker coefficient vector $\mathbf{w}_i$ is considered to include each speaker's information.
In other words, in this low-dimensional linear space spanned by the basis vectors $\mathbf{B}$, any speaker parameter vector can be represented by manipulating the speaker coefficient vector $\mathbf{w}_i$.
\mmedit{This paper refers to the low-dimensional model parameter linear space spanned by the basis vectors as speaker space.
Consequently, our proposed method can generate diverse speaker voices by varying the speaker coefficient vector $\mathbf{w}_i$ in the speaker space. Thus, this method is considered an eigenvoice approach. This indicates the applicability to speaker adaptation and speech style control along the dominant axis similar to the prior eigenvoice methods~\cite{eigenvoice_synthesis, eigenvoice1, eigenvoice2, eigenvoice_survey, eigenvoice_conversion}.}

\subsection{Sampling strategy of the speaker coefficient vector}
\label{subsection:sampling strategy}
To generate arbitrary speakers' voices, we use the sampling strategy for the speaker coefficient vector $\mathbf{w}$ that matches the statistics of the base speakers $\mathbf{a}_i, (i=1, \dots, N)$.
In this study, the speaker matrix $\mathbf{A}$ is standardized as a preprocessing step before performing SVD.
The statistic values for the seen base speaker parameter vectors are calculated as follows:
\begin{align}
\label{equation:sampling_method}
\mathbb{E}[\mathbf{a}_i] &=  \mathbf{0} \\
\mathrm{Cov}[\mathbf{a}_i] &= \frac{1}{N} \mathbf{A} \mathbf{A}^\top = \mathbf{U}\mathbf{\Sigma} \mathbf{V}^\top \mathbf{V} \mathbf{\Sigma}^\top \mathbf{U}^\top \\
&= \frac{1}{N} \mathbf{U} \mathbf{\Sigma} \mathbf{\Sigma}^\top \mathbf{U}^\top.
\end{align}
This study adopted a Gaussian sampling for a speaker coefficient vector $\mathbf{w}_n$ from a normal distribution $\mathcal{N}\left( \mathbf{0}, \frac{1}{N} \mathbf{I}\right)$. 
The statistical values of the arbitrary speaker parameter vector $\mathbf{a}_n$ are then formulated as bellow:
\begin{align}
& \mathbf{w}_n \sim \mathcal{N}\left(\mathbf{0}, \frac{1}{N}  \mathbf{I}\right) \\
\mathbb{E}[\mathbf{a}_n] &= \mathbf{U} \mathbf{\Sigma} E[\mathbf{w}_n] =  \mathbf{0} \\
\mathrm{Cov}[\mathbf{a}_n] &= \mathbf{U} \mathbf{\Sigma} \mathrm{Cov}[\mathbf{w}_n] \mathbf{\Sigma}^\top \mathbf{U}^\top = \frac{1}{N} \mathbf{U} \mathbf{\Sigma} \mathbf{\Sigma}^\top \mathbf{U}^\top.
\end{align}
Under this sampling strategy, these statistic values are consistent with those of the seen base speaker parameter vector $\mathbf{a}_i$.
Consequently, by sampling the speaker coefficient vector $\mathbf{w}_i$ from the normal distribution $\mathcal{N}\left( \mathbf{0}, \frac{1}{N} \mathbf{I}\right)$, we can generate arbitrary speaker parameter vectors $\mathbf{a}_n$ that statistically match the seen base speaker parameter vectors $\mathbf{a}_i$.

\begin{table}[t!]
    \begin{center}
    \caption{The MOS and WER score with 95\% confidence interval (CI) for the comparative method and proposed method.}
    \vspace{-5pt} %
    \scalebox{0.96}{
        \begin{tabular}{ cc|cc} 
            \toprule
            \multicolumn{2}{c}{Methods} & {MOS ($\uparrow$)} & {WER ($\downarrow$)} \\
            \toprule
            \toprule
             \multicolumn{2}{c|}{GT} & {4.29 $\pm$ 0.11}  & {4.20 $\pm$ 0.10} \\
             \toprule
             \multirow{2}{*}{x-vector} & base speaker &  {3.96 $\pm$ 0.13} &  {4.21 $\pm$ 0.09}\\
             & generated speaker & {3.99 $\pm$ 0.14} &  {2.51 $\pm$ 0.06}  \\
             \toprule
             \multirow{2}{*}{proposed} &  base speaker & {4.00 $\pm$ 0.12} &  {3.18 $\pm$ 0.06}\\
             & generated speaker & {3.81 $\pm$ 0.16} &  {2.20 $\pm$ 0.06}\\
            \bottomrule
        \end{tabular}
    }
    \label{table:mos_evaluation}
    \end{center}
          \vspace{-25pt} %
\end{table}

\section{Experiments}
\subsection{Experimental settings}

\subsubsection{Datasets}
\label{subsubsection:datasets}

In the experiments, we used the VCTK~\cite{vctk} dataset which contains approximately 44 hours of English speech from 109 speakers. From this dataset, we extracted 10 speakers' data (p225, p226, p227, p228, p229, p230, p231, p232, p237, and p241) including 5 male and 5 female speakers as the base speaker subset.

\subsubsection{Base Speaker Models}

We adopted the implementation and configuration of Conformer-FastSpeech2 (CFS2)~\cite{cfs2, conformer, fastspeech2} from ESPnet~\cite{espnet}. For waveform generation, we used a pre-trained HiFi-GAN~\cite{hifigan} vocoder from the ParallelWaveGAN repository\footnote{\url{https://github.com/kan-bayashi/ParallelWaveGAN}}.
As base speaker models, we created single-speaker TTS models for each of the 10 speakers by fine-tuning on their respective data.
Then, we used the same pre-trained model initialization trained on the full VCTK dataset.
During fine-tuning, we updated only the parameters of the variance adaptor and decoder module as these components are speaker-dependent.
In total, $5 \times 10^{10}$ model parameters were updated through fine-tuning.

\subsubsection{Comparative Method}
\label{subsection:comparative_method}

As a comparative method, we developed an x-vector-based multi-speaker TTS model using the same CFS2 architecture.
As a speaker encoder, this approach employed 
a pre-trained x-vector module~\cite{xvector} from SpeechBrain~\cite{speechbrain} (trained on the VoxCeleb1~\cite{voxceleb1} and VoxCeleb2~\cite{voxceleb2} datasets).
This baseline model was trained from scratch using the 10 base speakers’ data, as described in Section~\ref{subsubsection:datasets}. 
To generate new speakers' voices using the baseline model, we adopted the same sampling strategy as the proposed model. 
\mmedit{
Specifically, we applied SVD on the stack of the x-vectors and created new speaker x-vectors using the x-vector speaker space obtained by SVD.}

\begin{figure}[t]
   \centering
    \begin{minipage}[b]{1\linewidth}
      \centering
      \includegraphics[width=\hsize]{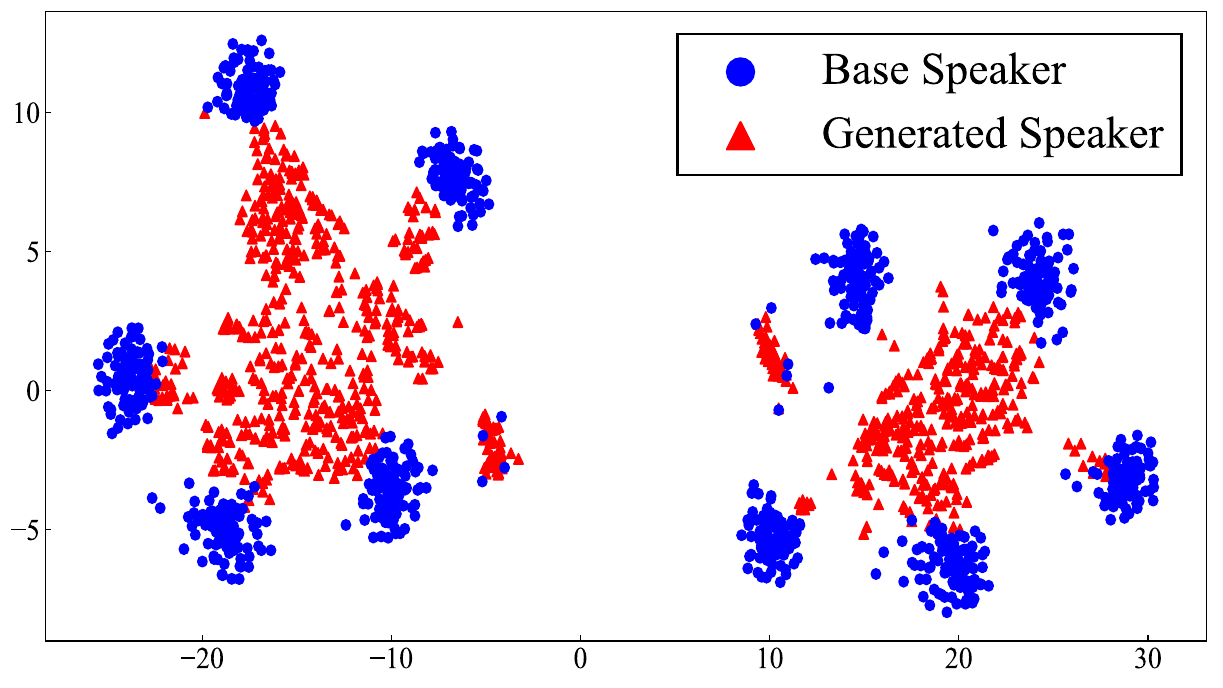}
    \end{minipage}
        \vspace{-20pt} %
    \caption{t-SNE visualization \mmedit{of the speaker embeddings of the base (blue dots) and generated speakers (red triangles).}}
    \vspace{-15pt} %
    \label{fig:t_sne}
  \end{figure}
\subsection{Speech quality evaluation}
\label{sec:mos_wer_test}

In speech synthesis, generated speech sometimes include artifacts that can lead to naturalness and intelligibility degradation. Thus, to evaluate naturalness and intelligibility, we employed the mean opinion score (MOS) and word error rate (WER) metrics, respectively.
We generated speech samples with 10 test sentences randomly selected from the VCTK test set for 100 generated speakers.
We also evaluated the ground truth speech (\textbf{GT}) from the base speakers in the VCTK dataset and synthesized speech from each base speaker TTS model (\textbf{base speaker}).
We denote our proposed method as ``\textbf{proposed}'' and the x-vector-based comparative method as ``\textbf{x-vector}''. 

For the MOS evaluation, 40 participants rated the naturalness of the given speech samples on a 5-point scale. Each participant evaluated 5 randomly selected speech samples for each method, resulting in a total of 25 speech samples per participant.
Participants were recruited through a crowdsourcing platform, Amazon Mechanical Turk\footnote{\url{https://www.mturk.com/}}.
For the intelligibility evaluation, we calculated WER scores using a pre-trained Conformer model~\cite{conformer} from ESPnet Model Zoo\footnote{\url{https://github.com/espnet/espnet_model_zoo}}.

Table~\ref{table:mos_evaluation} shows the MOS and WER results for each method.
Compared with the x-vector-based method, the proposed method achieved a comparable MOS score and a slightly better WER score.
\mmedit{Being consistent with previous research on model merge interpolation~\cite{merge_attribute}, this indicates that the proposed method can generate speech with high quality on par with the x-vector-based approach.
However, it must be noted that the primary focus of this work is not to achieve high-quality speech synthesis quality but to generate diverse speaker characteristics.}

\begin{figure}[t]
   \centering
    \begin{minipage}[b]{1\linewidth}
      \centering
      \includegraphics[width=\hsize]{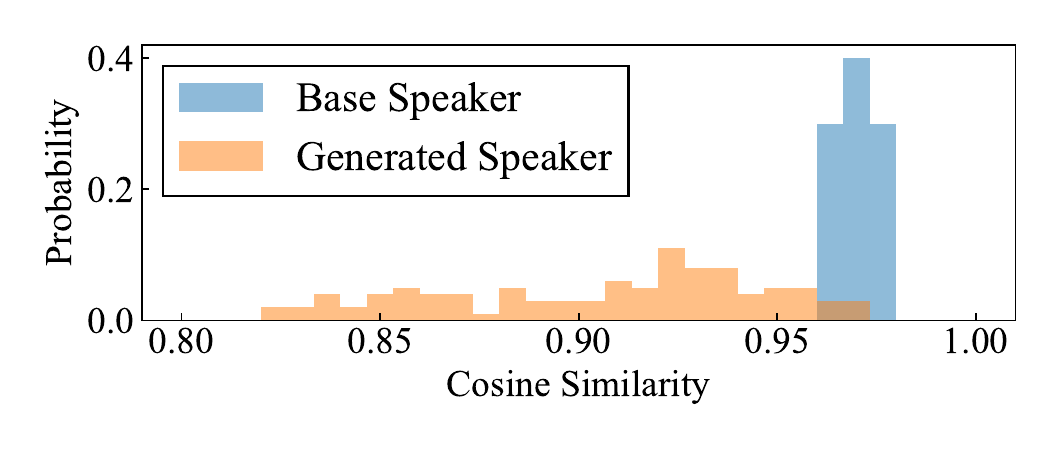}
    \end{minipage}
        \vspace{-25pt} %
    \caption{\mmedit{The distribution of the cosine similarity between each generated speaker and the nearest base speaker (GT).}}
    \vspace{-15pt} %
    \label{fig:cos_sim_distribution}
  \end{figure}

\subsection{Speaker diversity evaluation}
\label{sec:spk_diversity}

\mmedit{To evaluate the diversity of the generated speakers, we calculated the similarities of speaker embeddings of generated and base speakers.}
We generated speech samples using 10 test sentences for 100 generated speakers.
We also used the synthesized speech from the 10 base speaker TTS models with 100 test sentences randomly selected from the VCTK test set.
\mmedit{Then, we extracted speaker embeddings from the generated speakers and the base speakers (both GT speech and TTS-synthesized speech) using the Resemblyzer toolkit~\cite{resemblyzer}.}

We first investigated the distribution of speaker embeddings of base speakers (TTS) and generated speakers using t-SNE.
Figure~\ref{fig:t_sne} shows 
the t-SNE visualization of the speaker embeddings.
\mmedit{
The base speakers (blue dots) were clustered into distinct groups corresponding to the 10 base speakers, consisting of 5 male and 5 female speakers.
In contrast, the generated speakers (red triangles) were more scattered than those of the base speakers and surrounded by these base male and female clusters.
This suggests that the proposed method can generate diverse speaker voices, including both male and female.}

\mmedit{To evaluate the diversity of generated speakers, we also investigated the distribution of cosine similarities to the speaker embeddings of GT base speakers.}
\mmedit{
For each generated speaker, we calculated the similarity between the generated speaker and each of the 10 GT-based speakers.
Then, we obtained the maximum cosine similarity to the 10 GT speakers (i.e., the similarity to the most similar base speaker).
The same procedure was followed for the base speakers (TTS) and generated speakers (x-vector-based).}
\mmedit{If the maximum similarity was close to 1.0, it meant that the generated speaker was very similar to any of the base speakers. Otherwise, it can be said that the generated speaker had various characteristics that differed from the base speakers.}
Figure~\ref{fig:cos_sim_distribution} shows the distribution of maximum cosine similarities between the speaker embeddings of the generated speaker (proposed) and the GT base speakers. 
\mmedit{In our proposed method, the maximum cosine similarities were distributed from 0.82 to 0.97, indicating that generated speakers included speakers with different speaker characteristics from that of the base speakers. 
Even compared with the generated speaker from x-vector-based method, in which maximum cosine similarities were distributed from 0.78 to 0.97, our proposed method achieved comparable speaker diversity.
Thus, }the result showed that our proposed method is capable of representing a broader range of speaker characteristics compared to the base TTS models.%

\begin{figure}[t]
   \centering
    \begin{minipage}[b]{1\linewidth}
      \centering
      \includegraphics[width=\hsize]{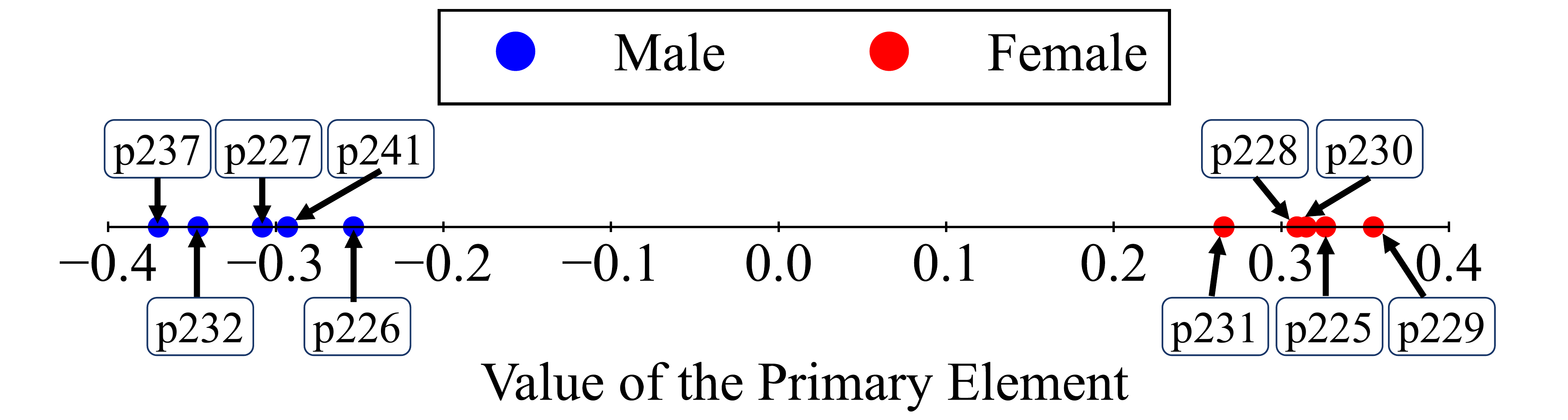}
    \end{minipage}
        \vspace{-20pt} %
    \caption{The visualization of the primary elements of speaker coefficient vectors of base speaker parameter vectors. \mmedit{Each dots correspond to the base speakers.}}
    \label{fig:cos_sim_matrix}
  \end{figure}

\begin{figure}[t]
  \centering
\includegraphics[width=\hsize]{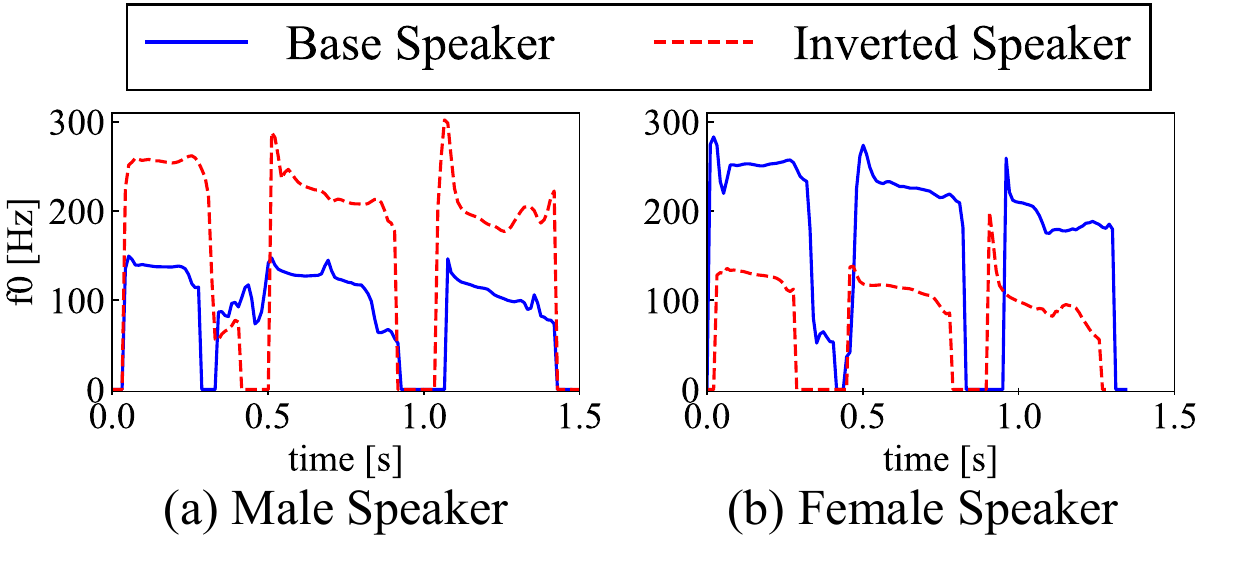}
  \vspace{-22pt}
  \caption{\mmedit{$F_0$ plot of the base speaker (blue solid) and inverted speaker (red dash) samples for both cases where the original speaker is male and female.}}
  \label{fig:f0_plot}
  \vspace{-15pt}
\end{figure}
\section{Discussion}

For further understanding of the extracted eigenvoice axis, we analyzed the base speaker coefficient vectors.
Through the analysis of the speaker coefficient vectors $\mathbf{w}_i , (i=1, \dots, N)$ obtained via SVD, we observed that the signs (positive or negative number) of the first principal component of the speaker coefficient vector corresponded to the gender attribute, as shown in Figure~\ref{fig:cos_sim_matrix}. 
To investigate this tendency, we conducted an experiment where we inverted only the sign of the first component of each 10-base speaker's coefficient vector and generated the corresponding inverted speaker parameter vector. The resulting models, based on the inverted speaker parameter vectors, generated speech with the opposite gender to the original one.
\mmedit{We also observed the pitch curve differences between the synthesized speech of the original speaker and the inverted speaker by using the Harvest~\cite{harvest} method.
In Figure~\ref{fig:f0_plot}, distinct pitch differences are observed between the original and inverted speakers for both cases where the original speaker is male and female.}
This indicates the possibility that the first basis axis of the speaker space is related to gender attributes and the possibility of controlling gender attributes by manipulating model parameters along this axis.

\section{Conclusions}
\label{sec:conclusion}

In this study, we \mmedit{proposed a model editing-based eigenvoice synthesis method to generate diverse speakers' voices.}
We \mmedit{conducted subjective and objective evaluations of the generated speakers' voices} from various perspectives.
The experimental results demonstrated that our method can generate diverse speakers’ speech while maintaining \mmedit{high naturalness and intelligibility.}
Furthermore, we discovered a gender-dominant attribute axis, such as gender, in the created speaker space, indicating the potential for attribute manipulation.

In future work, we plan to
explore various attribute spaces \mmedit{created by our method, such as emotions and styles.}

\bibliographystyle{IEEEtran}
\bibliography{mybib}

\end{document}